# 5G is Real: Evaluating the Compliance of the 3GPP 5G New Radio System with the ITU IMT-2020 Requirements

**Samer Henry, Ahmed Alsohaily, Member, IEEE, and Elvino Sousa, Fellow, IEEE**
Department of Electrical and Computer Engineering, University of Toronto, Toronto, ON M5S2E4, Canada

**ABSTRACT** The 3rd Generation Partnership Project (3GPP) submitted the 5G New Radio (NR) system specifications to International Telecommunication Union (ITU) as a candidate fifth generation (5G) mobile communication system (formally denoted as IMT-2020 systems). As part of the submission, 3GPP provided a self-evaluation for the compliance of 5G NR systems with the ITU defined IMT-2020 performance requirements. This paper considers the defined 5G use case families, Ultra Reliable Low-Latency Communication (URLLC), massive Machine Type Communication (mMTC) and enhanced Mobile Broadband (eMBB), and provides an independent evaluation of the compliance of the 3GPP 5G NR self-evaluation simulations with the IMT-2020 performance requirements for connection density, reliability, and spectral efficiency for future mobile broadband and emerging IoT applications. Independent evaluation indeed shows the compliance of the 3GPP 5G NR system with the ITU IMT-2020 performance requirements for all parameters evaluated by simulations.

## I. INTRODUCTION

Mobile communication applications have shifted from basic voice telephony to empowering a wide range of verticals across various industries, most notably via the rapidly expanding Internet of Things (IoT) applications, and are expected to continue to grow, occupy an integral part of our lives and ultimately transform societies as a whole [1], [12], [13]. While the fourth generation of mobile communication systems, formally referred to as International Mobile Telecommunications-Advance (IMT-Advanced) systems, provided a versatile platform for enabling a wide range of Mobile Broadband applications (and, to a certain extent, Low Power Wireless Internet of Things (IoT) applications [12]), the increasing potential for disruptive IoT applications with very high deployment densities (millions of devices in a relatively small areas) was one of the main motivators for the development of the next generation of mobile communication systems, the International Mobile Telecommunications-2020 (IMT-2020) commonly referred to as 5G – the fifth generation of mobile communication systems. The other motivators were increasing demand for enhanced mobile broadband services and the vast potential for mobile communications providing ultra-low latency and ultra-high reliability [1] – [3] for higher frequencies [24]. Candidate IMT-2020 systems are undergoing a rigorous evaluation process to ensure they fulfill the requirements set out by the International Telecommunication Union (ITU) for IMT-2020 systems, illustrated in Fig.1, to meet the performance requirements of emerging 5G applications, commonly grouped into enhanced Mobile Broadband (eMBB), Ultra-Reliable Low-Latency Communications (URLLC) and massive Machine Type Communications (mMTC) [1] – [5], [25]. The prime IMT-2020 candidate system, the 5G New Radio (NR) system developed by the 3rd Generation Partnership Project (3GPP) [6] – [10], promises to fulfill the IMT-2020 system requirements set out by the ITU [2] – [5] as detailed in the 3GPP self-evaluation submission [11]. Nevertheless, it is of utmost importance to independently verify the validity of the 3GPP submission prior to officially declaring the 5G NR system as an IMT-2020 compliant system.

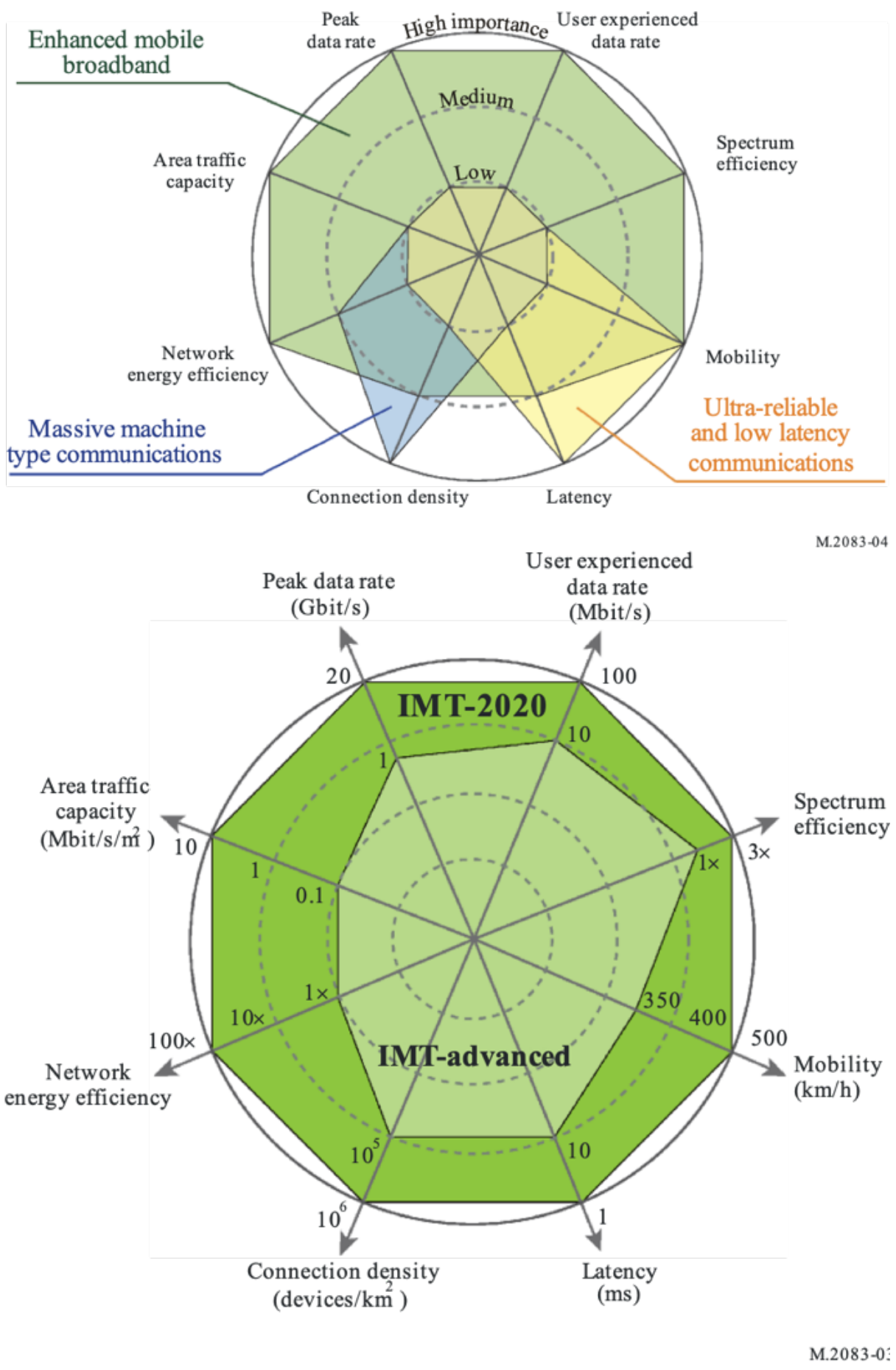

FIGURE 1. IMT-2020 use case scenarios (top) and performance requirements (bottom) (reproduced from [7])

This paper focuses on assessing the performance of the 3GPP 5G NR system for applications in the areas of massive Machine Type Communications (mMTC), Ultra-Reliable Low-Latency Communications (URLLC) and enhanced Mobile Broadband (eMBB), which are expected to play an integral role in future Internet of Things (IoT) applications, with focus on key parameters evaluated by system simulations to providing an independent evaluation to the compliance of the 3GPP submitted 5G NR self-evaluation simulations [11] via a custom simulator, which considered numerous academic and industrial simulations [14], [18]-[21] and compares the results of 3GPP developed simulations by companies such as Huawei, Ericsson, Intel and NTT Docomo among others. Some of these applications include, but are not limited to: smart wearables, health monitors, autonomous driving, and remote computing [25].
The contributions of this work are as follows: (i) a detailed system-level simulator for evaluating 5G candidate systems and (ii) an evaluation of the simulator performance in achieving 5G requirements for IMT-2020 in comparison with other industrial simulators for multiple test environments. The rest of this paper is as follows. Overviews of IMT-2020 system requirements, evaluation processes and scenarios are in Section II. The system structure for performance evaluation and additional features are detailed in Section III. Section IV discusses the system setup and methodology for simulation and the simulation results are detailed in Section V. Section VI concludes the paper and the appendix details tables providing requirements and results for each assessment as well as the results.

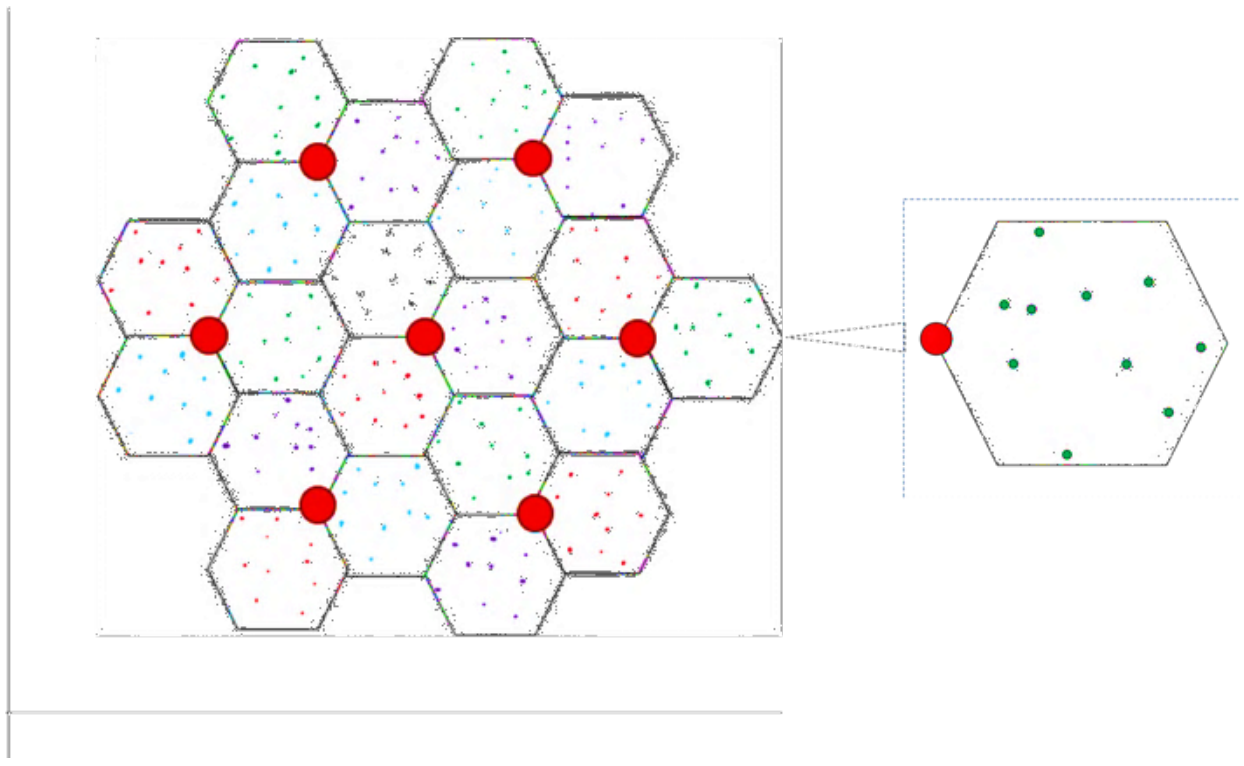

FIGURE 2. Urban Macro, Urban Micro, and Rural scenarios with UE distribution (UEs with the same color communicate with the same base station)

## II. SYSTEM REQUIREMENTS, EVALUATION PROCESS, AND SCENARIOS

### A. EVALUATION GUIDELINES

The simulator acts as an evaluation tool for the submitted 3GPP proposal [7] as per the specified evaluation methodology and configurations in the 3GPP report. System-level and link-level simulations are performed using our simulation tool to provide an independent evaluation of the 3GPP self-evaluation, which provides a complete compliance documentation for several technologies with the minimum IMT-2020 performance requirements.

### B. TEST ENVIRONMENTS

Five specific test environments are defined [22]-[23] for evaluating compliance with the performance requirements of IMT-2020 systems: Indoor hotspot-eMBB, Dense Urban-eMBB, Rural-eMBB, Urban Macro-mMTC, and Urban Macro-URLLC. Simulation of all test environments (with the exception of Indoor Hotspot-eMBB) uses a wrap-around configuration of 19 sites as shown in figures 2 – 4, each of 3TRxPs (cells) creating a hexagonal layout.
Antenna element distribution, cell range, and inter-site distance (ISD) is considered for geometry. The indoor hotspot scenario models a 120m x 50 m building floor with 12 Base stations placed 20 meters apart as per Figure 3. The Dense urban area consists of a macro layer following a 3-TRxP hexagonal layout, and a micro layer with 3 micro-sites randomly dropped in each TRxP area a number of user equipment (UE) distributed in the area. The rural eMBB test environment follows the macro layer of the dense urban area. A high-speed test environment is shown in figure 4 for mobility scenarios of UEs moving at 30 km/h, 120 km/h, and 500 km/h.

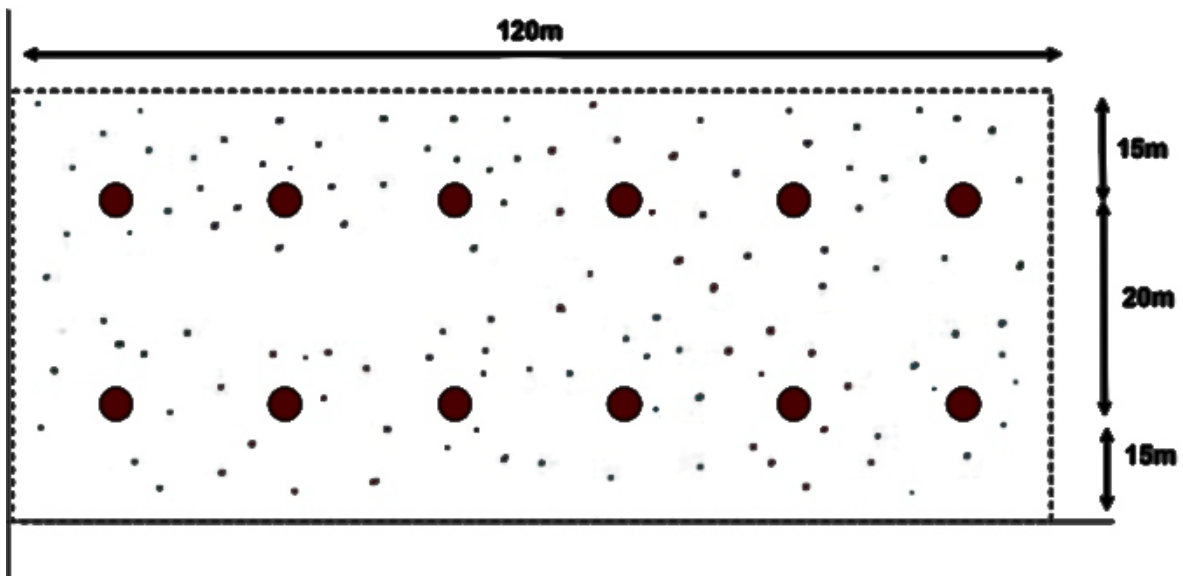

**FIGURE 3.** Indoor scenario with 12 access points (larger circles) and distributed users (smaller circles)

### C. EVALUATION CRITERIA

For evaluating system performance using simulations, the following key parameters are taken into consideration:

#### 1) SPECTRAL EFFICIENCY

The average spectral efficiency is obtained by running system-level simulations over a number of drops for each of the following three test environments: Indoor Hotspot-eMBB, Dense Urban-eMBB, and Rural-eMBB. Each drop is a sum of correctly received bits by all users over time as per the following equation [7]:

$$\widehat{SE}_{avg} = \frac{\sum_{j=1}^{N_{drops}} \sum_{i=1}^{N} R_i^{(j)}(T)}{N_{drops}\, T.W.M} \quad (1)$$

Where $\widehat{SE}_{avg}$ is the estimated average spectral efficiency that approaches the average by increasing the number of $N_{drops}$, $R_i^{(j)}(T)$ is the correct number of received bits during time $T$ for user $i$ in drop $j$, $W$ is the channel bandwidth, $N$ is the number of users, $M$ is the number of transmission/reception points between each transmit/receive antenna element pair. The 5th percentile user efficiency is the lowest 5th percentile point in the CDF of all users. The requirements for IMT2020 are detailed in Tables 1 and 2.

#### 2) CONNECTION DENSITY

The connection density is the total number of devices fulfilling a specific quality of service (QoS) per unit area (per km$^2$). The connection density is defined as [7]:

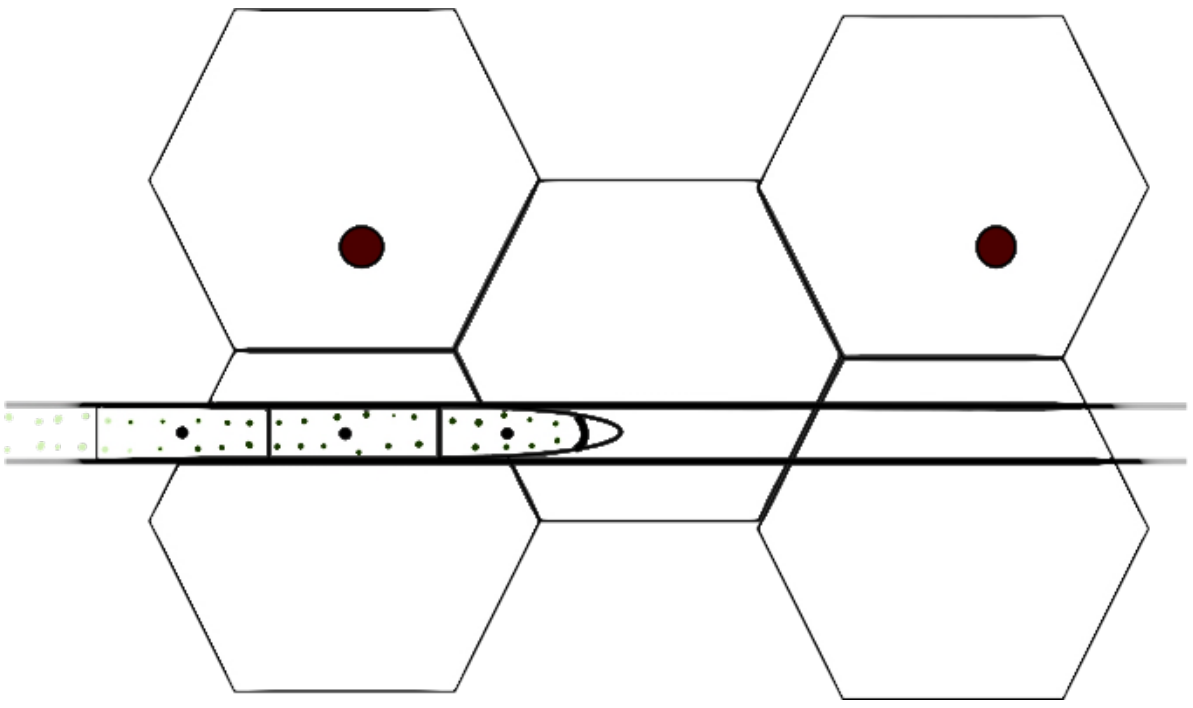

**FIGURE 4.** High speed mobility scenario (large circles are basestations, small circles are UEs)

$$C = \frac{N_{mux}.\, W/mean(B_i)}{ISD^2.\sqrt{3}/6} \quad (2)$$

Where $N_{mux}$ is the average number of multiplexed users for a given $SINR_i$, ISD is the Inter-site distance, and $B_i$ is as follows [7]:

$$B_i = T/(R_i/W_{user}) \quad (3)$$

The requirement is fulfilled if the 99th percentile of delay per user is less than or equal to 10 seconds and the system achieves a connection density of at least one million devices per square kilometer, evaluated for the Urban Macro scenario using simulations.

TABLE I
5TH PERCENTILE USER SPECTRAL EFFICIENCY REQUIREMENTS (REPRODUCED FROM [7])

| Test environment | Downlink (bit/s/Hz) | Uplink (bit/s/Hz) |
|---|---|---|
| Indoor Hotspot – eMBB | 0.3 | 0.21 |
| Dense Urban – eMBB (NOTE 1) | 0.225 | 0.15 |
| Rural – eMBB | 0.12 | 0.045 |

TABLE II
AVERAGE SPECTRAL EFFICIENCY REQUIREMENTS (REPRODUCED FROM [7])

| Test environment | Downlink (bit/s/Hz/TRxP) | Uplink (bit/s/Hz/TRxP) |
|---|---|---|
| Indoor Hotspot – eMBB | 9 | 6.75 |
| Dense Urban – eMBB | 7.8 | 5.4 |
| Rural – eMBB | 3.3 | 1.6 |

TABLE III
URLLC PERFORMANCE METRICS (REPRODUCED FROM [6])

| URLLC Performance Metric | Minimal Value |
|---|---|
| User plane latency | 1 ms (URLLC) |
| Control plane latency | (10 ms encouraged) |
| Reliability | 99.999% |
| Mobility Interruption time | 0 ms |

TABLE IV
ASSESSMENT METHODS FOR URLLC PERFORMANCE METRICS (REPRODUCED FROM [6])

| Characteristic for Evaluation | Assessment Method | Related Section of ITU-R Reports |
|---|---|---|
| User plane latency | Analytical | Report ITU-R M.[IMT-2020]. § 4.7.1 |
| Control plane latency | Analytical | Report ITU-R M.[IMT-2020]. § 4.7.2 |
| Reliability | Simulation | Report ITU-R M.[IMT-2020]. § 4.10 |
| Mobility Interruption time | Analytical | Report ITU-R M.[IMT-2020]. § 4.12 |

TABLE V
MOBILITY CLASSES (FROM [7])

| | Test environments for eMBB | | |
|---|---|---|---|
| | Indoor Hotspot – eMBB | Dense Urban – eMBB | Rural – eMBB |
| Mobility classes supported | Stationary, Pedestrian | Stationary, Pedestrian, Vehicular (up to 30 km/h) | Pedestrian, Vehicular, High speed vehicular |

### 3) RELIABILITY

Reliability is defined as the success probability (1- $P_e$) in which $P_e$ is the residual packet error ratio within maximum delay time as a function of SINR taking retransmission into account. The minimum requirement for the reliability is $1\text{-}10^{-5}$ success probability of transmitting a layer 2 PDU (protocol data unit) of 32 bytes within 1 ms in channel quality of coverage edge for the Urban Macro-URLLC test environment, assuming small application data (such as 20 bytes application data + protocol overhead). The requirement is fulfilled via downlink/uplink and LOS/NLOS as per Tables 3 and 4.

### 4) MOBILITY

Mobility is the maximum mobile station speed at which a defined QoS can be achieved (in km/h). The successful evaluation of mobility is to fulfill the threshold values for the packet error ratio and spectral efficiency for a mobility of 120km/h and 500 km/h. Table 5 defines the mobility classes that are to be supported in the respective test environments. A mobility class is supported if the traffic channel link data rate on the uplink, normalized by bandwidth, meets the criteria specified in Tables 5 and 6.

### 5) USER-EXPERIENCED DATA RATE

User experienced data rate is the 5% point of the cumulative distribution function (CDF) of the user throughput. The target values for the UE data rate are 100MBits/s for downlink and 50MBits/s for the uplink user experienced data rate.

TABLE VI
TRAFFIC CHANNEL LINK DATA RATE REQUIREMENTS NORMALIZED BY BANDWIDTH (FROM [7])

| Test environment | Normalized traffic channel link data rate (bit/s/Hz) | Mobility (km/h) |
|---|---|---|
| Indoor Hotspot – eMBB | 1.5 | 10 |
| Dense Urban – eMBB | 1.12 | 30 |
| Rural – eMBB | 0.8 | 120 |
| | 0.45 | 500 |

## III. SIMULATION STRUCTURE AND FEATURES

In this section, a description of our system level simulator structure and methodology are introduced for evaluating the requirements. Simulations are performed to evaluate each requirement independently with the exception of the joint evaluation of 5th percentile user spectral efficiency and the average spectral efficiency as simulations are performed to simultaneously evaluate them.

The simulator structure is entirely modular as shown in Figure 5 and supports multi-link transmissions. A spatial geometry application is integrated for single and multiple antenna configurations to obtain results. A Graphical User Interface (GUI) allows users to choose whether to set variables manually, choose from a predetermined test scenario, or optimize the placement of antenna elements by choosing an algorithm as per figure 6.

Using the GUI, values are assigned to parameters as per the user choice in the previous stage. Once again, value assignment can be predetermined or set manually. The number of drops and time durations set the complexity level for the loop in the next stage. Each time iteration, and once all parameters are defined, transmitters and receivers are deployed in two-dimensional or three-dimensional modes depending on the desired degree of complexity. Finally, the transmit/receive antenna configurations and antenna element patterns are defined. Figure 7 shows an example of choice of parameters. Simulations are then performed for all drops in which the SINR and performance is computed. Once the parameters are initialized, the system then loops the desired configuration scenarios. The inner loop calculates the performance for each transmit/receive antenna element pair, adding the following into consideration: interference, path loss, antenna gain (shown in figure 8), and antenna beam-steering properties. This is enclosed within another loop that combines the received signals between antennas for the time duration indicated during the input stage as maximum ratio combining or proportional fair scheduling. The third outer loop is to repeat the inner two loops for each user normally distributed around the environment space (either two or three-dimensional). The fourth outer loop repeats the simulation for the indicated number of drops for the results in section V, with an average of 10,000 drops are used. The Result Generation stage provides performance assessments, tables, and cumulative distribution functions of the SINR for considered test environments. The process is repeated until the iteration results converge as shown in figure 9.

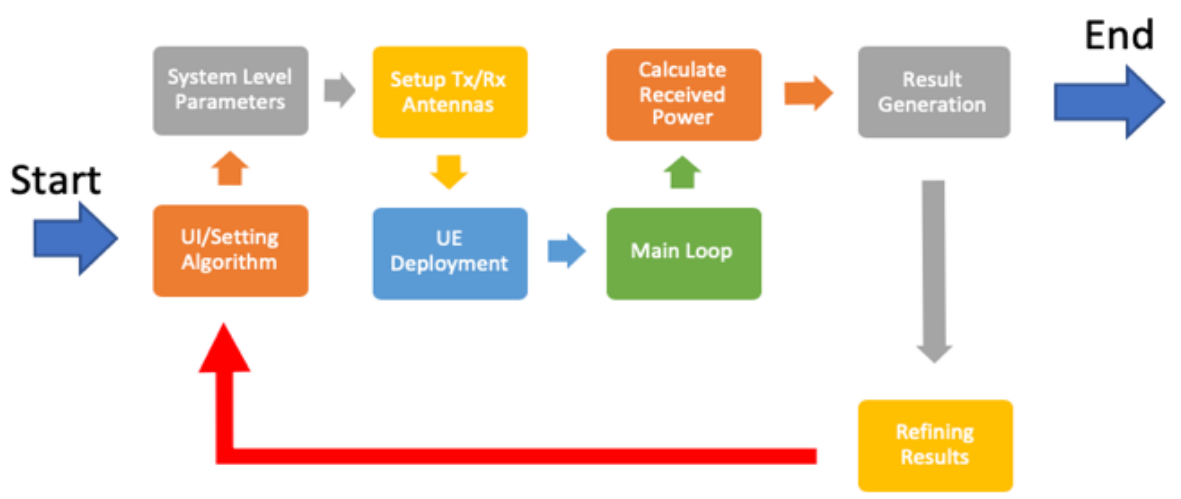

**FIGURE 5.** Modular Structure of the Simulator

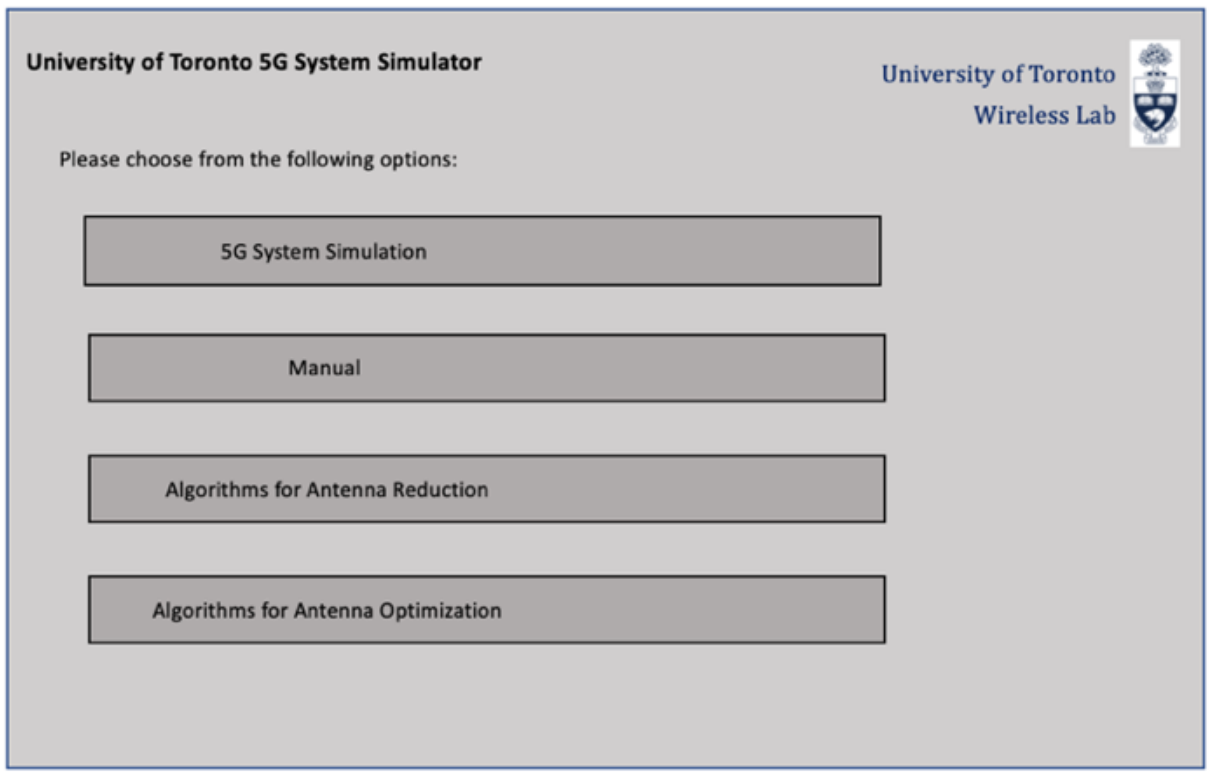

**FIGURE 6.** Simulator user Interface with mode options.

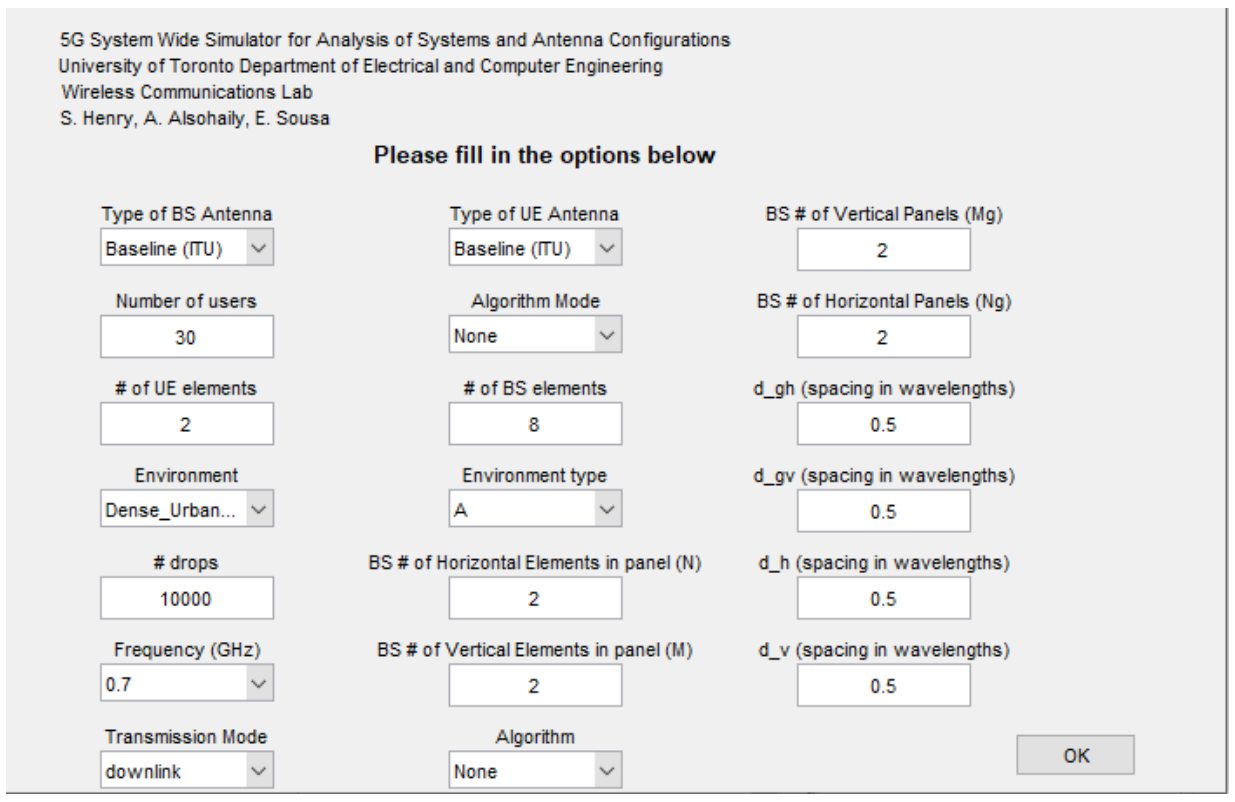

**FIGURE 7.** User-defined parameter menu.

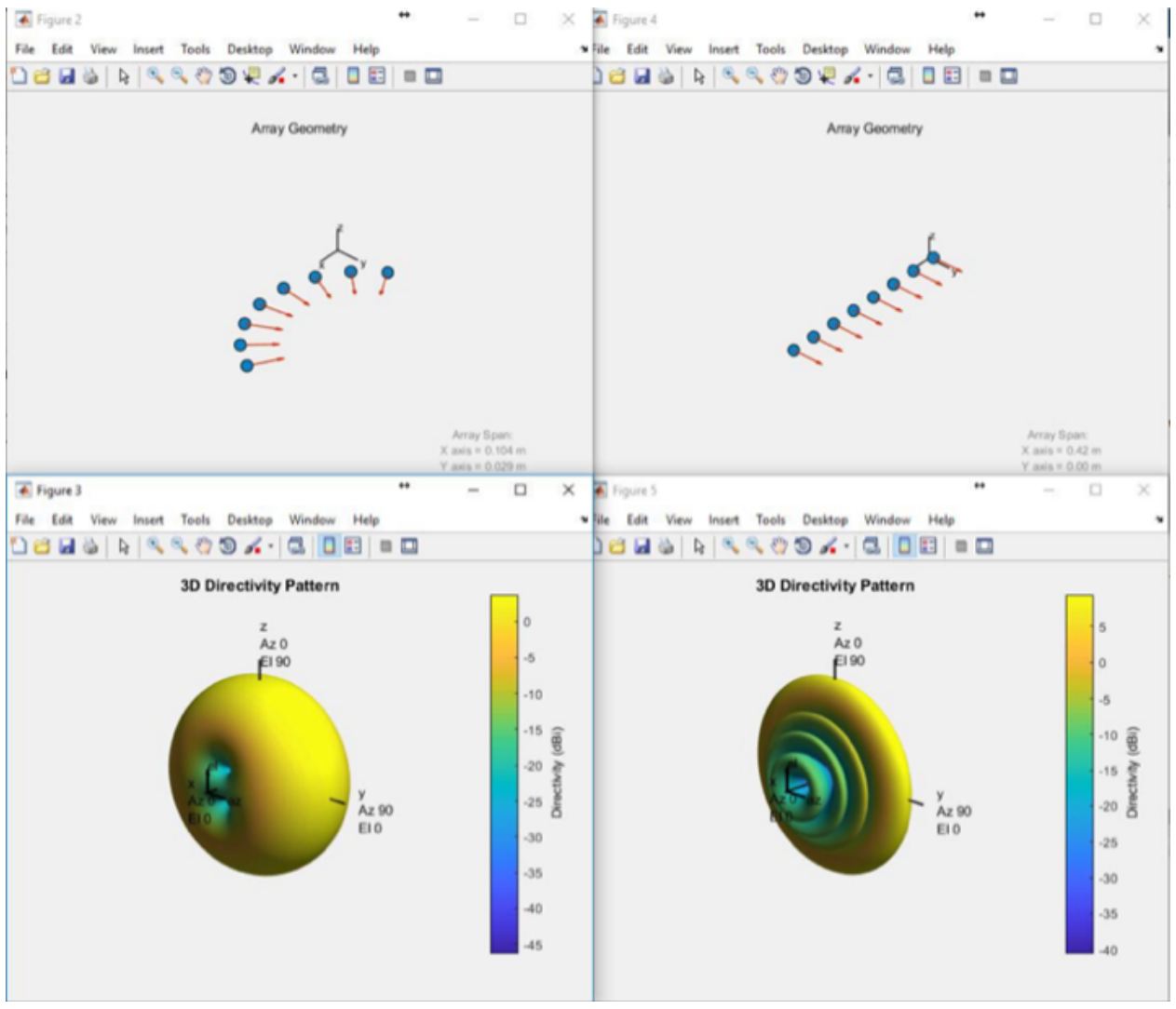

**FIGURE 8.** Antenna Gain and Directivity calculation via simulation.

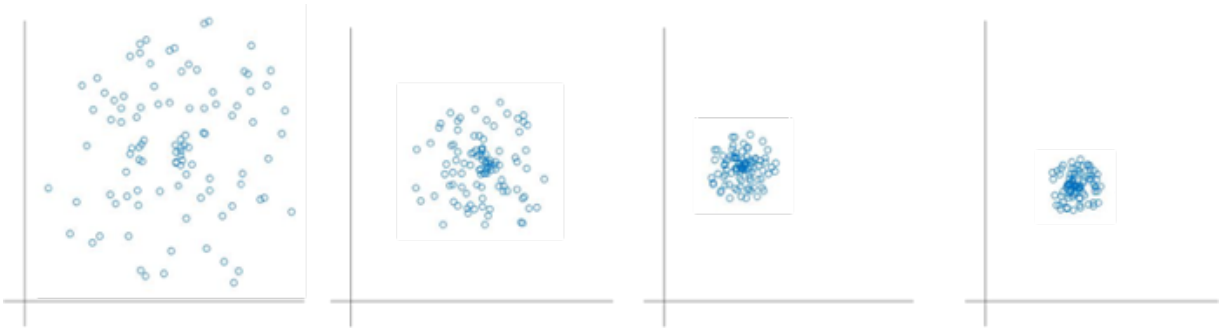

**FIGURE 9.** Convergence of SINR averages for URLLC calculations for 100 drops, 1000 drops, 5000 drops, and 10,000 drops (left to right).

## IV) SYSTEM SETUP AND SIMULATION METHODOLOGY

For the system-level simulation, user equipment (UE) are dropped independently over a predefined area of the network layout throughout the system and are modelled according to their respective traffic model. Each UE is randomly assigned LOS/NLOS channel conditions according to the channel model. Cell assignment to a UE is based on the cell selection scheme with applicable distances between UE and a base station depend on the proposed scenario. Signal fading and interference from each transmitter to each receiver is aggregated; interference over thermal parameter is taken into account as an uplink design constraint with an average interference of less than 10 dB. For full buffers, infinite queue depths are assumed. Channel quality, feedback delay, feedback errors, protocol data unit error which are inclusive of channel estimation error are modelled and packets are retransmitted according to the packet scheduler. For every drop, the simulation is run and repeated with UEs dropped at new random locations. 10,000 drops are performed for each simulation to ensure convergence in the system performance metrics of corresponding mean values. Finally, error modelling for channel estimation, phase noise, and control channels to decode the traffic channel is included.

## V) SIMULATION RESULTS

Based on the test environments and performance requirements outlined in Section II, simulations are performed using the simulator and methodology described in Sections III and IV. The tables and figures provided in this section detail the simulation results for the 3GPP 5G NR system and compare them to the ITU IMT-2020 requirements. The results indeed show the compliance of the 3GPP 5G NR system with the ITU IMT-2020 performance requirements for all parameters evaluated by simulations.

### A. CONNECTION DENSITY SIMULATION RESULTS

Taking into account layers 1 and 2 overhead information provided by the proponents, the connection density requirement is fullfilled if it is greater than the ITU report in [11] as shown in Tables 7-10. These four tables compare full-buffer and non full-buffer modes, scenarios A and B, and base-station inter-site distances of 1732 m and 500 m for system-level simulations between the University of Toronto, Huawei, and Ericsson simulators. The tables show that full-buffer outperforms non full-buffer for NB-IoT, mMTC, and NR technologies, and are compliant with ITU requirements.

TABLE VII

REQUIREMENTS FOR MMTC CONNECTION DENSITY, 500 M, NON-FULLBUFFER, SCENARIO A

| RIT | Antenna config & Tx scheme | Procedure | Numerology | Req. | | Huawei | Ericsson | Univ of Toronto |
|---|---|---|---|---|---|---|---|---|
| FDD | | | | | | | | |
| NB-IoT | 1x2 SIMO, Single-tone | Early data transmission | 15 kHz SCS | Connection density (/km2) | 1000000 | 8,047,087 | | 6,381,492 |
| | | | | Bandwidth (kHz) | | 180 | | 180 |
| NB-IoT | 1x2 SIMO, (15-180 kHz) | RRC Resume, data after Msg5, RRC Connection Release | 15 kHz SCS | Connection density (/km2) | 1000000 | | 1,233,000 | 1,190,895 |
| | | | | Bandwidth (kHz) | | | 180 | 180 |
| eMTC (LTE-M) | 1x2 SIMO | RRC Resume, data after Msg5, RRC Connection Release | 15 kHz SCS | Connection density (/km2) | 1000000 | | 1,893,000 | 1,489,022 |
| | | | | Bandwidth (kHz) | | | 360 | 360.00 |

### B. CONNECTION DENSITY CDF

In addition to the connection density values, figure 10 displays the cumuative distribution function of the aforementioned technologies in the previous section and the higher-then-average uplink SINR of the University of Toronto simulator compared to other industry simulators.

TABLE VIII

REQUIREMENTS FOR MMTC CONNECTION DENSITY, 1732 M, NON-FULLBUFFER, SCENARIO A

| RIT | Antenna config & Tx scheme | Procedure | Numerology | Req. | | Huawei | Ericsson | Univ of Toronto |
|---|---|---|---|---|---|---|---|---|
| FDD | | | | | | | | |
| NB-IoT | 1x2 SIMO, Single-tone | Early data transmission | 15 kHz SCS | Connection density (/km2) | 1000000 | 1,198,000 | | 1,074,221 |
| | | | | Bandwidth (kHz) | | 360 | | 360 |
| NB-IoT | 1x2 SIMO,(15-180 kHz) | RRC Resume, data after Msg5, RRC Connection Release | 15 kHz SCS | Connection density (/km2) | 1000000 | | 1,018,000 | 1,004,539 |
| | | | | Bandwidth (kHz) | | | 2700 | 2700 |
| eMTC (LTE-M) | 1x2 SIMO, DFT-S-OFDM | Early data transmission | 15 kHz SCS | Connection density (/km2) | 1000000 | 1,107,000 | | 1,055,847 |
| | | | | Bandwidth (kHz) | | 540 | | 540 |
| eMTC (LTE-M) | 1x2 SIMO, DFT-S-OFDM | RRC Resume, data after Msg5, RRC Connection Release | 15 kHz SCS | Connection density (/km2) | 1000000 | | 1,026,000 | 1,009,807 |
| | | | | Bandwidth (kHz) | | | 3240 | 3240 |

TABLE IX

REQUIREMENTS FOR MMTC CONNECTION DENSITY, 1732 M, NON-FULLBUFFER, SCENARIO A

| RIT | Antenna config & Tx scheme | Numerology | Req. | | Huawei | Ericsson | Intel | Univ of Toronto |
|---|---|---|---|---|---|---|---|---|
| FDD | | | | | | | | |
| NB-IoT | 1x2 SIMO, Single-tone (16Rx@BS) | 15 kHz SCS | Connection density (/km2) | 1000000 | 41,325,000 | 46,058,578 | | 35927461 |
| | | | Bandwidth (kHz) | | 180 | 180 | | 180 |
| eMTC | 1x2 SIMO, DFT-S-OFDM (16Rx@BS) | 15 kHz SCS | Connection density (/km2) | 1000000 | 39,995,000 | 25,674,701 | 40,036,848 | 32,850,912 |
| | | | Bandwidth (kHz) | | 180 | 180 | 180 | 180 |
| NR | 1x2 SIMO, OFDMA (16Rx@BS) | 15 kHz SCS | Connection density (/km2) | 1000000 | 36,575,000 | 30,066,283 | 40,066,168 | 31,582,109 |
| | | | Bandwidth (kHz) | | 180 | 180 | 180 | 180 |

TABLE X

REQUIREMENTS FOR MMTC CONNECTION DENSITY, 1732 M, NON-FULLBUFFER, SCENARIO A

| RIT | Antenna config & Tx scheme | Numerology | Req. | | Huawei | Ericsson | Intel | ZTE | Univ of Toronto |
|---|---|---|---|---|---|---|---|---|---|
| FDD | | | | | | | | | |
| NB-IoT | 1x2 SIMO, Single-tone (16 Rx@BS) | 15 kHz SCS | Connection density (/km2) | 1000000 | 2,517,000 | 2,237,326 | | 2,251,631 | 2,314,259 |
| | | | Bandwidth (kHz) | | 180 | 180 | | 180 | 180.00 |
| eMTC | 1x2 SIMO, DFT-S-OFDM (16 Rx@BS) | 15 kHz SCS | Connection density (/km2) | 1000000 | 1,344,000 | 1,231,947 | 1,062,780 | | 1,214,735 |
| | | | Bandwidth (kHz) | | 180 | 180 | 180 | | 180.00 |
| NR | 1x2 SIMO, OFDMA (16 Rx@BS) | 15 kHz SCS | Connection density (/km2) | 1000000 | 1,138,000 | 1,269,767 | 1,237,402 | 1,424,456 | 1,267,312 |
| | | | Bandwidth (kHz) | | 180 | 180 | 180 | 180 | 180.00 |

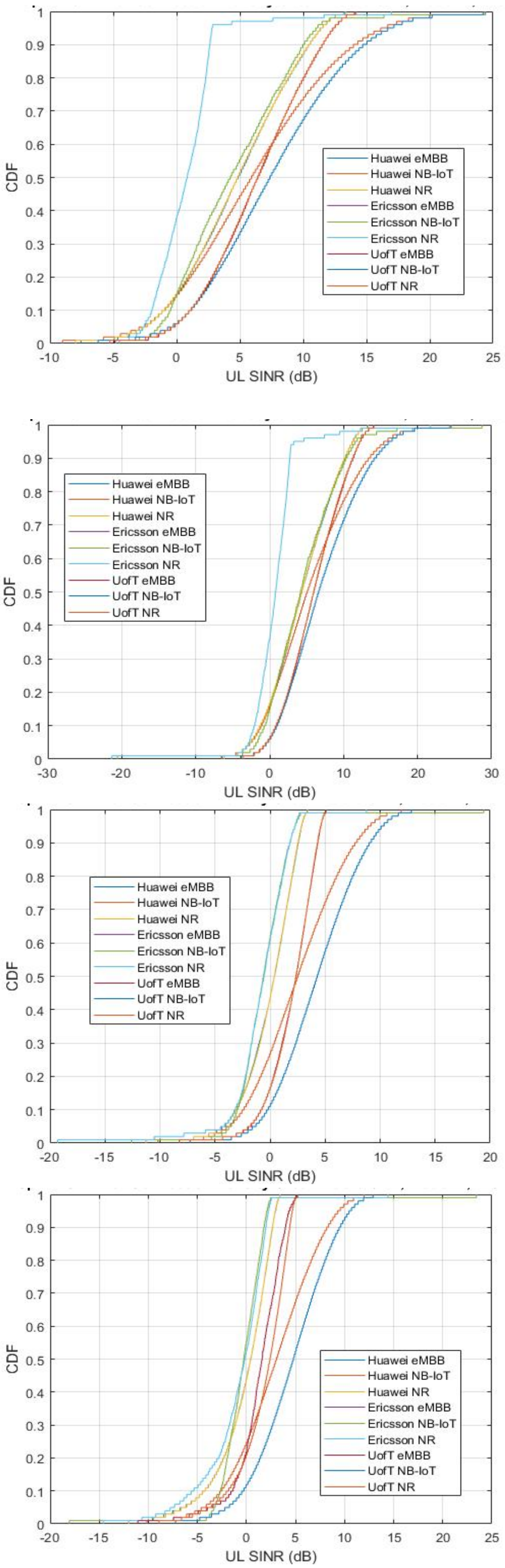

FIGURE 10. CDF of Uplink SINR, Connection Density, Dense Urban Test Scenario for Model A 500 m - Model B 1732 m - Model A 500 m - Model B 1732 m (top to bottom)

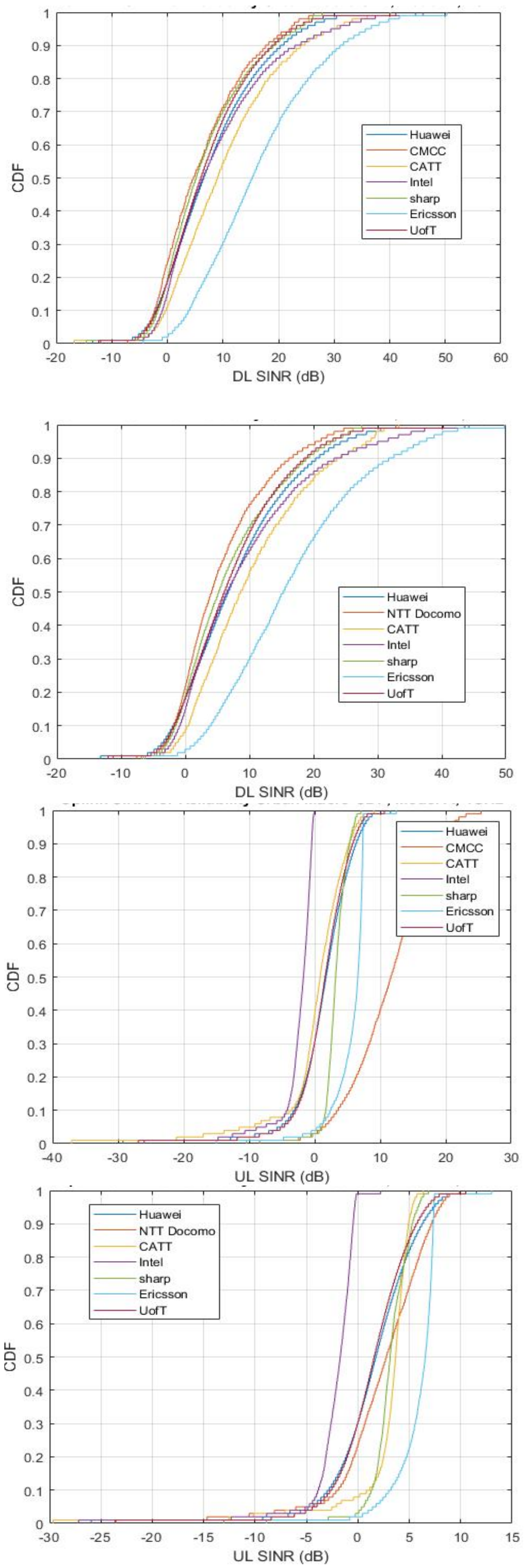

FIGURE 11. CDF of SINR, Reliability, Urban Macro Test Scenario at 4GHz for Downlink Model A - Downlink Model B - Uplink Model A - Uplink Model B (top to bottom)

### C. RELIABILITY SIMULATION RESULTS

Ultra-high reliability and good resilience capability are needed to achieve the reliability requirement for ensuring the 5th percentile downlink or uplink value within the required delay obtains a success probability equal to or higher than the required success probaility. Figure 11 and Table 11 both display the Uplink SINR for a 4 GHz spectrum and reliability results for 700MHz/4GHz respectively for 5-7 evaluators, and our simulator hence achieves the reliability requirements (>99.999%) as well as exceeding all testing scenarios and antenna configurations.

TABLE XI

COMPARING RELIABILITY RESULTS FOR URLLC CONFIGURATION

f=700 MHz

| Channel | Antenna Configuration | URLLC Requirement | Huawei | CATT | Nokia | Intel | Univ of Toronto |
|---|---|---|---|---|---|---|---|
| Channel Model A Downlink Reliability | 2x2 SU-MIMO 14os one-shot (1 PDCCH+1 PDSCH) | 99.999% | 99.9998% | | | | 99.99969% |
| | 2x2 SU-MIMO 7os one-shot (1 PDCCH+1 PDSCH) | 99.999% | 99.9998% | 99.99904% | | > 99.9999% | 99.9997% |
| | 2x2 SU-MIMO 4os slot aggregation = 2 (1 PDCCH + 2 PDSCH) | 99.999% | 99.9995% | | | | 99.999263% |
| | 16x4 SU-MIMO 14os (1PDCCH + 1 PDSCH) | 99.999% | | | 99.999% | | |
| Channel Model A Uplink Reliability | 1x8 SIMO, OFDMA, 14os one-shot (1 PUSCH) | 99.999% | 99.9999% | | | | 99.9999% |
| | 1x2 SIMO, OFDMA, 7os one-shot (2 PUSCH) | 99.999% | | | 9.9992% | | |
| | 1x8 SIMO 7os (1 PUSCH) | 99.999% | | | | > 99.9999% | |
| | 1x8 SIMO, OFDMA, 14os one-shot (1 PUSCH, 8RB) | 99.999% | | 99.9993% | | | |
| Channel Model B Downlink Reliability | 2x2 SU-MIMO 14os one-shot (1 PDCCH+1 PDSCH) | 99.999% | 99.9998% | 99.99904400% | | 99.99969% | |
| | 2x2 SU-MIMO 7os one-shot (1 PDCCH+1 PDSCH) | 99.999% | 99.999% | 99.9991% | > 99.9999% | 99.9998% | |
| | 2x2 SU-MIMO 4os slot aggregation = 2 (1 PDCCH + 2 PDSCH) | 99.999% | 99.9995% | | | | |
| Channel Model B Uplink Reliability | 1x8 SIMO, OFDMA, 14os one-shot (1 PUSCH) | 99.999% | 99.9999% | | | 99.9999941% | |
| | 1x2 SIMO, OFDMA, 7os one-shot (2 PUSCH) | 99.999% | | | | | |
| | 1x8 SIMO 7os (1 PUSCH) | 99.999% | | | > 99.9999% | | |
| | 1x8 SIMO, OFDMA, 14os one-shot (1 PUSCH, 8RB) | 99.999% | | 99.999% | | | |

f= 4 GHz

| Channel | Antenna Configuration | URLLC Requirement | Huawei | Nokia | Intel | Univ of Toronto |
|---|---|---|---|---|---|---|
| Channel Model A Downlink Reliability | 2x2 SU-MIMO 14os slot aggregation = 2 (1 PDCCH + 2 PDSCH) | 99.999% | 99.9998% | | | 99.9999% |
| | 32x8 SU-MIMO 14os (1PDCCH + 1 PDSCH) | 99.999% | | 99.999% | | |
| | 2x4 SU-MIMO 7os slot aggregation = 1 (1 PDCCH + 1 PDSCH) | 99.999% | | | > 99.9999% | |
| Channel Model A Uplink Reliability | 1x16 SIMO, OFDMA | 99.999% | 99.9998% | | | 99.9997% |
| | 1x8 SIMO 7os (1 PUSCH) | 99.999% | | | > 99.9999% | |
| | 1x2 SIMO 4os (2 PUSCH) | 99.999% | | 99.999% | | |
| Channel Model B Downlink Reliability | 2x2 SU-MIMO 14os slot aggregation = 2 (1 PDCCH + 2 PDSCH) | 99.999% | 99.99990561% | | 99.99969% | |
| | 32x8 SU-MIMO 14os (1PDCCH + 1 PDSCH) | 99.999% | | | | |
| | 2x4 SU-MIMO 7os slot aggregation = 1 (1 PDCCH + 1 PDSCH) | 99.999% | | > 99.9999% | | |
| | 2x2 SU-MIMO 4os (2 PDCCH + 2 PDSCH) | 99.999% | | | | 99.99997% |
| Channel Model B Uplink Reliability | 1x16 SIMO, OFDMA, | 99.999% | 99.9999% | | 99.9999% | 99.99993% |
| | 1x8 SIMO 14os (1 PUSCH, 12RB) | 99.999% | | > 99.9999% | | |

### D. SPECTRAL EFFICIENCY

Enhanced spectral efficiency results are included in Tables 12-15 for Indoor hotspot, dense urban, and rural evaluation scenarios for different TRxP and simulation bandwidths. Using the evaluation configuration parameters, the results show the data conforms with reference values and industry evaluators.

TABLE XII

SIMULATION RESULTS FOR DOWNLINK SPECTRAL EFFICIENCY, INDOOR HOTSPOT, 12 TRXP, 4 GHZ, CHANNEL MODEL A

| RIT | Antenna and TXRU mapping | Antenna config & Tx scheme | Numerology | Req. | | Huawei | Intel | Samsung | Univ of Toronto |
|---|---|---|---|---|---|---|---|---|---|
| **FDD** | | | | | | | | | |
| NR | gNB: (M,N,P,Mg,Ng; Mp,Np) = (4,4,2,1,1;4,4) | 32x4 MU-MIMO Type II Codebook | 15 kHz SCS | Average [bit/s/Hz/TRxP] | 9 | 11.287 | 10.627 | 13.160 | 9.812 |
| | | | | 5th percentile [bit/s/Hz] | 0.3 | 0.356 | 0.398 | 0.330 | 0.359 |
| **TDD** | | | | | | | | | |
| NR | gNB: (M,N,P,Mg,Ng; Mp,Np) = (4,4,2,1,1;4,4) | 32x4 MU-MIMO, 4T SRS | 30 kHz SCS | Average [bit/s/Hz/TRxP] | 9 | 12.965 | | | 11.122 |
| | | | | 5th percentile [bit/s/Hz] | 0.3 | 0.377 | | | 0.324 |
| | | | | | | | | | |
| NR | gNB: (M,N,P,Mg,Ng; Mp,Np) = (4,4,2,1,1;4,4) | 32x4 MU-MIMO, 4T SRS | 15 kHz SCS | Average [bit/s/Hz/TRxP] | 9 | 12.773 | | | 10.109 |
| | | | | 5th percentile [bit/s/Hz] | 0.3 | 0.394 | | | 0.381 |

TABLE XIII

SIMULATION RESULTS FOR DOWNLINK SPECTRAL EFFICIENCY, INDOOR HOTSPOT, 36 TRXP, 4 GHZ, CHANNEL MODEL A

| RIT | Antenna and TXRU mapping | Antenna config & Tx scheme | Numerology | Frame structure | Req. | | Huawei | Intel | Univ of Toronto |
|---|---|---|---|---|---|---|---|---|---|
| **FDD** | | | | | | | | | |
| NR | gNB: (M,N,P,Mg,Ng; Mp,Np) = (8,16,2,1,1;2,8) | 32x4 MU-MIMO Type II Codebook (256Tx@gNB) | 15 kHz SCS | | Average [bit/s/Hz/TRxP] | 9 | 13.340 | 12.586 | 12.84 |
| | | | | | 5th percentile [bit/s/Hz] | 0.3 | 0.312 | 0.406 | 0.302 |
| **TDD** | | | | | | | | | |
| NR | gNB: (M,N,P,Mg,Ng; Mp,Np) = (8,16,2,1,1;2,8) | 32x4 MU-MIMO, 4T SRS (256Tx@gNB) | 30 kHz SCS | DDDSU | Average [bit/s/Hz/TRxP] | 9 | 14.218 | | 12.507 |
| | | | | | 5th percentile [bit/s/Hz] | 0.3 | 0.350 | | 0.3286 |
| | | | | | | | | | |
| NR | gNB: (M,N,P,Mg,Ng; Mp,Np) = (8,16,2,1,1;2,8) | 32x4 MU-MIMO, 4T SRS (256Tx@gNB) | 15 kHz SCS | DDDSU | Average [bit/s/Hz/TRxP] | 9 | 14.563 | | 11.86992 |
| | | | | | 5th percentile [bit/s/Hz] | 0.3 | 0.385 | | 0.3574 |

TABLE XIV

SIMULATION RESULTS FOR DOWNLINK SPECTRAL EFFICIENCY, INDOOR HOTSPOT, 12 TRXP, 30 GHZ, CHANNEL MODEL A/B

| RIT | Antenna and TXRU mapping | Antenna config & Tx scheme | Numerology | Req. | | Huawei | Intel | Univ of Toronto |
|---|---|---|---|---|---|---|---|---|
| **TDD** | | | | | | | | |
| NR | gNB: (M,N,P,Mg,Ng; Mp,Np) = (4,4,2,1,1;4,4); UE: (M,N,P,Mg,Ng; Mp,Np) = (2,4,2,1,2; 1,2) | 32x8 MU-MIMO, 4T SRS (2 panel@UE) | 60 kHz SCS | Average [bit/s/Hz/TRxP] | 9 | 11.599 | | 10.8511 |
| | | | | 5th percentile [bit/s/Hz] | 0.3 | 0.308 | | 0.3241 |

### E. MOBILITY

5G systems support low to high mobility applications and much enhanced data rates in accordance with user and service demands in multiple user environments. Figures 12-14 and Tables 16-21 exhibit the uplink SINR and the normalized channel link data rate for NLOS/LOS conditions under various spectrum bandwidths.

### F. UE DATA RATE

Coupled with NR usage scenario, Table 22 illustrates the data rate for different antenna configurations for uplink and downlink, showing multi-band macro layer data rates are greater than that of the single-band macro layer, hence fulfilling the Data Rate requirement of 100 Mbit/s (downlink) and 50 Mbits/s (uplink).

TABLE XV

SIMULATION RESULTS FOR DOWNLINK SPECTRAL EFFICIENCY, DENSE URBAN, EMBB, 4 GHZ, CHANNEL MODEL A

| Channel model A | RIT | Antenna and TXRU mapping | Antenna config & Tx scheme | Req. | | Huawei | Univ of Toronto |
|---|---|---|---|---|---|---|---|
| DL | FDD | | FDD | | | | |
| | NR | gNB: (M,N,P,Mg,Ng; Mp,Np) = (8,4,2,1,1;1,4) | 8x2 MU-MIMO Type II Codebook | Average [bit/s/Hz/TRxP] | 3.3 | 7.040 | 6.594 |
| | | | | 5th percentile [bit/s/Hz] | 0.12 | 0.180 | 0.138 |
| | TDD | | TDD | | | | |
| | NR | gNB: (M,N,P,Mg,Ng; Mp,Np) = (8,4,2,1,1;1,4) | 8x2 MU-MIMO, 2T SRS | Average [bit/s/Hz/TRxP] | 3.3 | 7.490 | 7.817 |
| | | | | 5th percentile [bit/s/Hz] | 0.12 | 0.159 | 0.193 |
| UL | FDD | | FDD | | | | |
| | NR | gNB: (M,N,P,Mg,Ng; Mp,Np) = (8,4,2,1,1;1,4) | 1x8 SU-MIMO, OFDMA | Average [bit/s/Hz/TRxP] | 1.6 | 3.718 | 4.279 |
| | | | | 5th percentile [bit/s/Hz] | 0.045 | 0.127 | 0.138 |
| | NR | gNB: (M,N,P,Mg,Ng; Mp,Np) = (8,4,2,1,1;1,4) | 2x8 SU-MIMO, Codebook based, OFDMA | Average [bit/s/Hz/TRxP] | 1.6 | 4.296 | 4.751 |
| | | | | 5th percentile [bit/s/Hz] | 0.045 | 0.093 | 0.104 |

| | | | | | | | | | |
|---|---|---|---|---|---|---|---|---|---|
| | FDD | | | | | | | | |
| | NR | gNB: (M,N,P,Mg,Ng; Mp,Np) = (8,8,2,1,1;2,8) | 32x4 MU-MIMO Type II Codebook (128Tx@gNB) | 15 kHz SCS | Average [bit/s/Hz/TRxP] | 7.8 | 11.450 | | 10.891 |
| | | | | | 5th percentile [bit/s/Hz] | 0.225 | 0.376 | | 0.358 |
| | | | | | | | | | |
| | TDD | | | | | | | | |
| | NR | gNB: (M,N,P,Mg,Ng; Mp,Np) = (8,8,2,1,1;2,8) | 32x4 MU-MIMO, 4T SRS (128Tx@gNB) | 30 kHz SCS | Average [bit/s/Hz/TRxP] | 7.8 | 13.042 | | 12.263 |
| | | | | | 5th percentile [bit/s/Hz] | 0.225 | 0.382 | | 0.361 |
| | | | | | | | | | |
| | NR | gNB: (M,N,P,Mg,Ng; Mp,Np) = (8,8,2,1,1;2,8) | 32x4 MU-MIMO, 4T SRS (128Tx@gNB) | 15 kHz SCS | Average [bit/s/Hz/TRxP] | 7.8 | 12.951 | | 11.886 |
| | | | | | 5th percentile [bit/s/Hz] | 0.225 | 0.388 | | 0.314 |
| | NR | gNB: (M,N,P,Mg,Ng; Mp,Np) = (12,8,2,1,1;4,8) | 64x4 MU-MIMO, 4T SRS (192Tx@gNB) | 30 kHz SCS | Average [bit/s/Hz/TRxP] | 7.8 | 16.098 | | 15.020 |
| | | | | | 5th percentile [bit/s/Hz] | 0.225 | 0.494 | | 0.462 |
| | | | | | | | | | |
| | NR | gNB: (M,N,P,Mg,Ng; Mp,Np) = (12,8,2,1,1;4,8) | 64x4 MU-MIMO, 4T SRS (192Tx@gNB) | 15 kHz SCS | Average [bit/s/Hz/TRxP] | 7.8 | 15.708 | | 14.809 |
| | | | | | 5th percentile [bit/s/Hz] | 0.225 | 0.484 | | 0.454 |

TABLE XVI

SIMULATION RESULTS FOR EMBB INDOOR HOTSPOT UPLINK MOBILITY, 4 GHZ, 12 TRXP, CHANNEL MODEL A

| RIT | Antenna config & Tx scheme | Numerology | Frame structure | Req. | | Channel cond. | Huawei | Univ of Toronto |
|---|---|---|---|---|---|---|---|---|
| FDD | | | | | | | | |
| NR | 1x8 SIMO, OFDMA | 15 kHz SCS | | Normalized traffic channel link data rate (bit/s/Hz) | 1.5 | NLOS | 1.75 | 1.63 |
| | | | | | 1.5 | LOS | 2.05 | 2.1 |
| TDD | | | | | | | | |
| NR | 1x8 SIMO, OFDMA | 30 kHz SCS | DDDSU | Normalized traffic channel link data rate (bit/s/Hz) | 1.5 | NLOS | 1.59 | 1.38 |
| | | | | | 1.5 | LOS | 1.94 | 1.82 |

TABLE XVII

SIMULATION RESULTS FOR EMBB INDOOR HOTSPOT UPLINK MOBILITY, 30 GHZ, 12TRXP, CHANNEL MODEL A/B

| RIT | Antenna config & Tx scheme | Numerology | Frame structure | Req. | | Channel cond. | NTT DOCOMO | Intel (SNR margin) | Samsung | Univ of Toronto |
|---|---|---|---|---|---|---|---|---|---|---|
| FDD | | | | | | | | | | |
| NR | 1x4 SIMO | 60kHz | | Normalized traffic channel link data rate (bit/s/Hz) | 1.5 | NLOS | 2.84 | | | 2.71 |
| | | | | | 1.5 | LOS | | | | |

TABLE XVIII

SIMULATION RESULTS FOR EMBB INDOOR HOTSPOT UPLINK MOBILITY, 4 GHZ, 36 TRXP, CHANNEL MODEL A

| RIT | Antenna config & Tx scheme | Numerology | Frame structure | Req. | | Channel cond. | Intel (SNR margin) | Univ of Toronto (SNR margin) |
|---|---|---|---|---|---|---|---|---|
| FDD | | | | | | | | |
| NR | 2x8 SIMO | 15kHz SCS | | | | NLOS | 0.39 | 0.34 |
| | | 30kHz SCS | | | | NLOS | 0.38 | 0.39 |

TABLE XIX

SIMULATION RESULTS FOR EMBB RURAL UPLINK MOBILITY, 700 MHZ, 120 KM/H CHANNEL MODEL A

| RIT | Antenna config & Tx scheme | Numerology | Frame structure | Req. | | Channel cond. | Huawei | Ericsson | Intel (SNR margin) | Univ of Toronto |
|---|---|---|---|---|---|---|---|---|---|---|
| FDD | | | | | | | | | | |
| NR | 1x4 SIMO, OFDMA | 15 kHz SCS | | Normalized traffic channel link data rate (bit/s/Hz) | 0.8 | NLOS | 2.32 | | | 2.13 |
| | | | | | 0.8 | LOS | 2.90 | | | 2.57 |
| TDD | | | | | | | | | | |
| NR | 1x4 SIMO, OFDMA | 15 kHz SCS | DDDSU | Normalized traffic channel link data rate (bit/s/Hz) | 0.8 | NLOS | 2.10 | | | 1.92 |
| | | | | | 0.8 | LOS | 2.63 | | | 2.18 |

TABLE XX

SIMULATION RESULTS FOR EMBB DENSE URBAN UPLINK MOBILITY, 4 GHZ, 12 TRXP, CHANNEL MODEL A

| RIT | Antenna config & Tx scheme | Numerology | Req. | | Channel cond. | Huawei | Ericsson | Univ of Toronto |
|---|---|---|---|---|---|---|---|---|
| FDD | | | | | | | | |
| NR | 1x8 SU-MIMO, OFDMA | 15 kHz SCS | Normalized traffic channel link data rate (bit/s/Hz) | 1.12 | NLOS | 1.92 | | 2.03 |
| | | | | 1.12 | LOS | 2.22 | | 2.29 |
| TDD | | | | | | | | |
| NR | 1x8 SU-MIMO, OFDMA | 30 kHz SCS | Normalized traffic channel link data rate (bit/s/Hz) | 1.12 | NLOS | 1.82 | | 1.76 |
| | | | | 1.12 | LOS | 2.17 | | 1.95 |

TABLE XXI

SIMULATION RESULTS FOR EMBB DENSE URBAN UPLINK MOBILITY, 700 MHZ, 500 KM/H, CHANNEL MODEL A

| RIT | Antenna config & Tx scheme | Numerology | Frame structure | Req | Channel cond. | Intel (SNR margin) | Univ of Toronto |
|---|---|---|---|---|---|---|---|
| FDD | | | | | | | |
| NR | 2x2 SIMO | 15kHz SCS | | | NLOS | | 1.28 |
| | | | | | LOS | | |

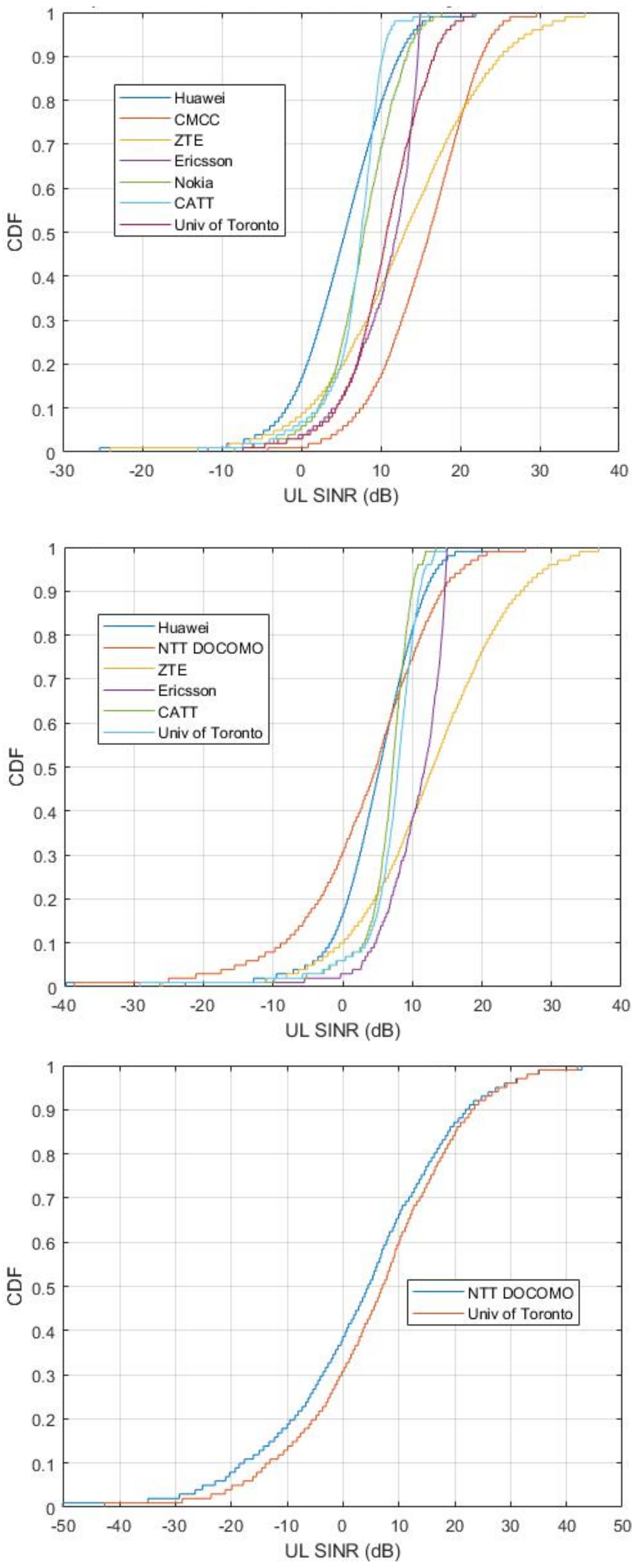

**FIGURE 12.** CDF of Uplink SINR, Mobility, Dense Urban Test Scenario at 4GHz for 4 GHz Model A – 4 Ghz Model B– 30 GHz Model A/B (top to bottom)

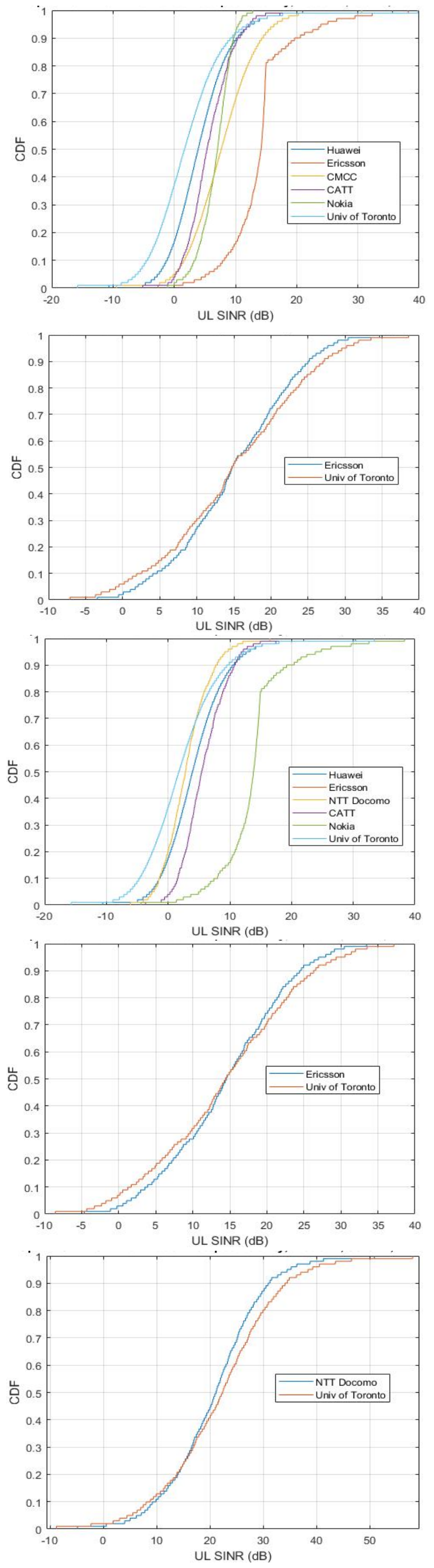

**FIGURE 13.** CDF of Uplink SINR, Mobility, Indoor Hotspot Test Scenario for 4 GHz, Model A, 12 TRxP – 4 GHz, Model B, 12 TRxP - 4 GHz, Model A, 36 TRxP – 4 GHz, Model B, 36 TRxP- 30 GHz, Model A, 12 TRxP (top to bottom)

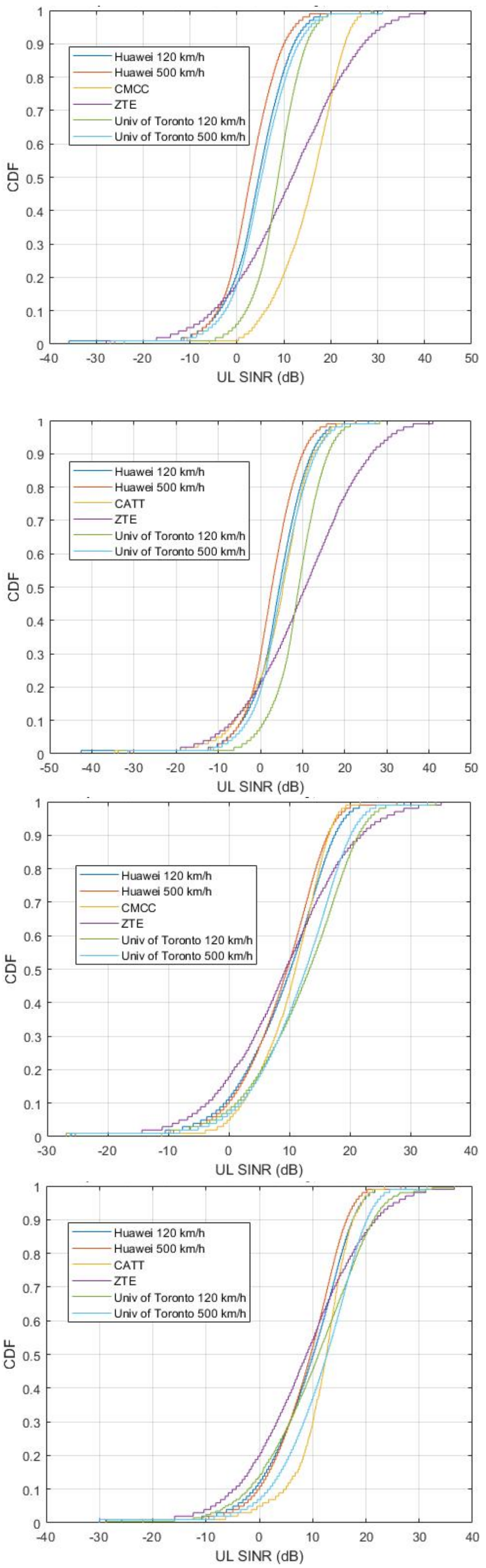

**FIGURE 14.** CDF of Uplink SINR, Mobility, Rural Test Scenario for 700 MHz, Model A– 700 MHz, Model B – 4 GHz, Model A– 4 GHz, Model B (top to bottom)

TABLE XXII

SIMULATION RESULTS FOR USER EXPERIENCE DATA, MODEL A, DOWNLINK/UPLINK RESPECTIVELY

| RIT | Detailed config. For multi-band/layer | Antenna config & Tx scheme | Numerology | Req. | | Huawei | Univ of Toronto |
|---|---|---|---|---|---|---|---|
| **Multi-band Macro layer** | | | | | | | |
| NR | Macro layer only 4 GHz | 32x4 MU-MIMO, CB based OFDMA | 15 kHz SCS | User experienced data rate (Mbit/s) | 100 | | 152.482 |
| **Single-band Macro layer** | | | | | | | |
| NR | Macro layer only 4 GHz | 32x4 MU-MIMO, CB based OFDMA | 15 kHz SCS | User experienced data rate (Mbit/s) | 100 | | 132.561 |
| **Multi-band Macro layer** | | | | | | | |
| NR | Macro layer only 30 GHz (TDD) + 4 GHz (SUL) 50% offload | 4 GHz (SUL band): 2x32 SU-MIMO, CB based OFDMA 30 GHz (TDD band): 8x32 SU-MIMO, CB based, OFDMA (2 panel@UE) | 4 GHz: 15 kHz SCS 30 GHz: 60 kHz SCS | User experienced data rate (Mbit/s) | 50 | 52.11 | 62.9103 |
| **Single-band Macro layer** | | | | | | | |
| NR | Macro layer only 4 GHz | 16x16 SU-MIMO, CB based OFDMA | 15 kHz SCS | User experienced data rate (Mbit/s) | 50 | | 59.3013 |
| NR | Macro layer only 30 GHz | 32x32 SU-MIMO, CB based OFDMA (2 panel@UE) | 120 kHz SCS | User experienced data rate (Mbit/s) | 50 | | 72.4814 |

## VI) SUMMARY, CONCLUSIONS, AND FUTURE WORK

This paper utilized an independent simulator to assess the compliance of the 3GPP submitted 5G NR self-evaluation simulations with the ITU IMT-2020 performance requirements. The results indeed confirm the compliance of the 3GPP 5G system with the ITU connection density, reliability, and mobility requirements to support the anticipated 5G applications and use cases. Building on this work, additional simulations can be performed for a wide range of frequency ranges and system configurations (rural, highway, etc.) to determine performance gaps and potential areas for improvement for the 3GPP 5G NR system.

## VIII) APPENDIX

### A. CONNECTION DENSITY PARAMETERS

**mMTC Simulation Algorithm:**

Loop over number of drops:
- Distribute users over environment
- Loop over users:
  - 1-Set environment, network layout, and antenna array parameters
  - 2-Assign propagation condition (LOS/NLOS)
  - 3-Calculate pathloss
  - 4-Generate large scale parameters
  - 5-Generate delays
  - 6-Generate cluster powers
  - 7-Generate arrival angles and departure angles for both azimuth and elevation
  - 8-Couple of rays within a cluster for both azimuth and elevation
  - 9-Generate the cross polarization power ratios
  - 10-Draw initial random phases
  - 11-Generate channel coefficients
  - 12-Apply pathloss and shadowing

Calculate Connection Density and SINR

**FIGURE A.1.** Procedure algorithm for evaluating

TABLE A.1
PARAMETERS FOR EVALUATING MMTC CONNECTION DENSITY
(REPRODUCED FROM [5])

| Connection Density Evaluation (Full-Buffer) | | |
|---|---|---|
| | **Configuration A** | **Configuration B** |
| Thermal noise level | -174 dBm/Hz | -174 dBm/Hz |
| Traffic model | With layer 2 PDU (Protocol Data Unit) message size of 32 bytes:<br>1 message/day/device or 1 message/2 hours/device<br>Packet arrival follows Poisson arrival process for non-full buffer system-level simulation | With layer 2 PDU (Protocol Data Unit) message size of 32 bytes:<br>1 message/day/device or 1 message/2 hours/device<br>Packet arrival follows Poisson arrival process for non-full buffer system-level simulation |
| Simulation bandwidth | Up to 10 MHz | Up to 50 MHz |
| UE density | Not applicable for non-full buffer system-level simulation as evaluation methodology of connection density<br>For full buffer system-level simulation followed by link-level simulation, 10 UEs per TRxP NOTE – this is used for SINR CDF distribution derivation | Not applicable for non-full buffer system-level simulation as evaluation methodology of connection density<br>For full buffer system-level simulation followed by link-level simulation, 10 UEs per TRxP NOTE – this is used for SINR CDF distribution derivation |
| UE antenna height | 1.5 m | 1.5 m |
| Carrier frequency for evaluation | 700 MHz | 700 MHz |
| BS antenna height | 25 m | 25 m |
| Total transmit power per TRxP | 49 dBm for 20 MHz bandwidth<br>46 dBm for 10 MHz bandwidth | 49 dBm for 20 MHz bandwidth<br>46 dBm for 10 MHz bandwidth |
| UE power class | 23 dBm | 23 dBm |
| Percentage of high loss and low loss building type | 20% high loss, 80% low loss | 20% high loss, 80% low loss |
| **Additional parameters for system-level simulation** | | |
| Inter-site distance | 500 m | 1732 m |
| Number of antenna elements per TRxP[1] | Up to 64 Tx/Rx | Up to 64 Tx/Rx |
| Number of UE antenna elements | Up to 2Tx/Rx | Up to 2Tx/Rx |
| Device deployment | 80% indoor, 20% outdoor<br>Randomly and uniformly distributed over the area | 80% indoor, 20% outdoor<br>Randomly and uniformly distributed over the area |
| UE mobility model | Fixed and identical speed \|v\| of all UEs, randomly and uniformly distributed direction | Fixed and identical speed \|v\| of all UEs, randomly and uniformly distributed direction |
| UE speeds of interest | 3 km/h for indoor and outdoor | 3 km/h for indoor and outdoor |
| Inter-site interference modelling | Explicitly modelled | Explicitly modelled |
| BS noise figure | 5 dB | 5 dB |
| UE noise figure | 7 dB | 7 dB |
| BS antenna element gain | 8 dBi | 8 dBi |
| UE antenna element gain | 0 dBi | 0 dBi |

### B. RELIABILITY REQUIREMENTS AND PARAMETERS

TABLE B.1
URLLC PERFORMANCE METRICS (REPRODUCED FROM [4])

| URLLC Performance Metric | Minimal Value |
|---|---|
| User plane latency (URLLC) | 1 ms (URLLC) |
| Control plane latency (URLLC) | (10 ms encouraged) |
| Reliability (URLLC) | 99.999% |
| Mobility interruption time (URLLC) | 0 ms |

TABLE B.2
ASSESSMENT METHODS FOR URLLC PERFORMANCE (REPRODUCED FROM [5])

| Characteristic for evaluation | Assessment method | Related section of ITU-R Reports |
|---|---|---|
| User plane latency | Analytical | Report ITU-R [IMT 2020.TECH PERF REQ], § 4.7.1 |
| Control plane latency | Analytical | Report ITU-R [IMT 2020.TECH PERF REQ], § 4.7.2 |
| Reliability | Simulation | Report ITU-R [IMT 2020.TECH PERF REQ], § 4.10 |
| Mobility interruption time | Analytical | Report ITU-R [IMT 2020.TECH PERF REQ], § 4.12 |

**URLLC Simulator Algorithm**

Loop over number of drops:
  Distribute users over environment
  Loop over users:
    1-Set environment, network layout, and antenna array parameters
    2-Assign propagation condition (LOS/NLOS)
    3-Calculate pathloss
    4-Generate large scale parameters
    5-Generate delays
    6-Generate cluster powers
    7-Generate arrival angles and departure angles for both azimuth and elevation
    8-Couple of rays within a cluster for both azimuth and elevation
    9-Generate the cross polarization power ratios
    10-Draw initial random phases
    11-Generate channel coefficients
    12-Apply pathloss and shadowing
Calculate Reliability and SINR

**FIGURE B.1.** Procedure algorithm for evaluating reliability for URLLC scenarios

TABLE B.3
PARAMETERS FOR EVALUATING URLLC RELIABILITY (REPRODUCED FROM [5])

| Parameters | Urban Macro–URLLC | |
|---|---|---|
| | Reliability Evaluation | |
| | Configuration A | Configuration B |
| Baseline evaluation configuration parameters | | |
| Carrier frequency for evaluation | 4 GHz | 700 MHz |
| BS antenna height | 25 m | 25 m |
| Total transmit power per TRxP | 49 dBm for 20 MHz bandwidth<br>46 dBm for 10 MHz bandwidth | 49 dBm for 20 MHz bandwidth<br>46 dBm for 10 MHz bandwidth |
| UE power class | 23 dBm | 23 dBm |
| Percentage of high loss and low loss building type | 100% low loss | 100% low loss |
| Additional parameters for system-level simulation | | |
| Inter-site distance | 500 m | 500 m |
| Number of antenna elements per TRxP[1] | Up to 256Tx/Rx | Up to 64 Tx/Rx |
| Number of UE antenna elements | Up to 8Tx/Rx | Up to 4Tx/Rx |
| Device deployment | 80% outdoor,<br>20% indoor | 80% outdoor,<br>20% indoor |
| UE mobility model | Fixed and identical speed \|v\| of all UEs, randomly and uniformly distributed direction | Fixed and identical speed \|v\| of all UEs, randomly and uniformly distributed direction |
| UE speeds of interest | 3 km/h for indoor and 30 km/h for outdoor | 3 km/h for indoor and 30 km/h for outdoor |
| Inter-site interference modelling | Explicitly modelled | Explicitly modelled |
| BS noise figure | 5 dB | 5 dB |
| UE noise figure | 7 dB | 7 dB |
| BS antenna element gain | 8 dBi | 8 dBi |
| UE antenna element gain | 0 dBi | 0 dBi |
| Thermal noise level | -174 dBm/Hz | -174 dBm/Hz |
| Traffic model | Full buffer<br>Note: This is used for SINR CDF distribution derivation | Full buffer<br>Note: This is used for SINR CDF distribution derivation |
| Simulation bandwidth | Up to 100 MHz<br>Note: This value is used for SINR CDF distribution derivation | Up to 40 MHz<br>Note: This value is used for SINR CDF distribution derivation |
| UE density | 10 UEs per TRxP<br>Note: This is used for SINR CDF distribution derivation | 10 UEs per TRxP<br>Note: This is used for SINR CDF distribution derivation |
| UE antenna height | 1.5 m | 1.5 m |
| Evaluated service profiles | Full buffer best effort | Full buffer best effort |
| Simulation bandwidth | Up to 100 MHz (for carrier frequency of 4 GHz) | Up to 40 MHz (for carrier frequency of 700 MHz) |

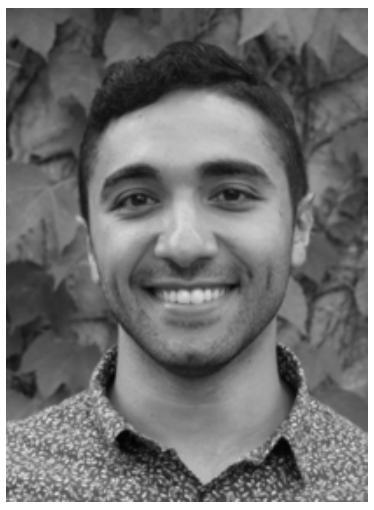

**SAMER HENRY** (M'15) received the B.Eng. degree (Hons.) in electrical engineering from the American University in Cairo in 2010, and the M.A.Sc degree from the University of Toronto in 2015, where he is currently pursuing the Ph.D. degree with the Department of Electrical and Computer Engineering. He is a Fulbright scholar as well as a German Exchange (DAAD) scholarship winner. He led a team in IBM to create a Highly Available Cluster Multiprocessing Protocol at IBM in 2012 and is currently a lecturer at the University of Toronto.

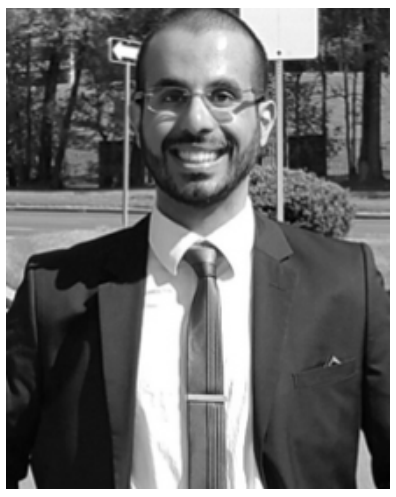

**AHMED ALSOHAILY** (M'15) received the B.Eng. degree (Hons.) in electrical engineering from King Saud University in 2010, and the M.Eng. and Ph.D. degrees from the University of Toronto in 2011 and 2015, respectively. He is currently a member of the Technology Strategy Team with TELUS, and a Post-Doctoral Fellow with the Department of Electrical and Computer Engineering, University of Toronto. He is also an Advisor for the Next Generation Mobile Networks Alliance, and actively contributes to the Third Generation Partnership Project and the IEEE Communication Society Standards Development.

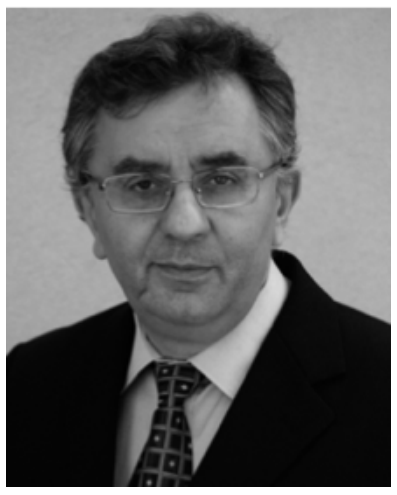

**ELVINO S. SOUSA** (S'79–M'80–SM'96–F'12) received the B.A.Sc. degree in engineering science and the M.A.Sc. degree in electrical engineering from the University of Toronto, in 1980 and 1982, respectively, (S'79–M'80–SM'96–F'12) received the B.A.Sc. degree in engineering science and the M.A.Sc. degree in electrical engineering from the University of Toronto, in 1980 and 1982, respectively, and the Ph.D. degree in electrical engineering from the University of Southern California, in 1985. He pioneered the area of wire- less communications with the University of Toronto, where he is currently the Director of the Wireless Laboratory, which has undertaken research on wireless systems for the past 28 years. Since 1986, he has been with the Department of Electrical and Computer Engineering, University of Toronto, where he is a Professor and the Jeffrey Skoll Professor of Computer Network Architecture. He has been invited to give numerous lectures and short courses on spread spectrum, code division multiple access, and wireless systems in many countries, and has been a Consultant to the industry and Governments internationally in the area of wireless systems. He has also been involved in various standardization and industry related wireless activities and is actively participating in NGMN as an Advisor. He is the Inventor of the autonomous infrastructure wireless network concept. His current interests are in the areas of autonomous infrastructure wireless networks, cognitive radio, self-configurable wireless networks, and two-tier networks. He was the Technical Program Chair of PIMRC'95, the Vice Technical Program Chair of Globecom'99, and the Co-Technical Program Chair of WPMC'10 and PIMRC'11. He has been involved in the technical program committee of numerous international conferences. He was the Chair of the IEEE Technical Committee on Personal Communications. He has spent sabbatical leaves with Qualcomm and Sony CSL/ATL. He has received the Queen Elizabeth II Golden Jubilee Medal.